\documentclass[aps,prd,11pt,showpacs,superscriptaddress,amssymb,amsmath,nofootinbib]{revtex4-1}

\usepackage{amsmath,amssymb}
\usepackage{graphicx}
\usepackage{xcolor}
\usepackage{multirow,booktabs}
\usepackage{tabularx}
\usepackage{array}
\usepackage{rotating}
\usepackage{hyperref}

\begin{document}

\title{Eccentricity-Modulated Phase Degeneracy and Distinguishability between Dark Matter and Accretion Disk Environmental Effects in EMRIs}

\author{Tian-Hao Wu}
\email{1020241999@glut.edu.cn}

\author{Shu-Jun Rong}
\email{rongshj@glut.edu.cn}

\affiliation{
College of Physics and Electronic Information Engineering,
Guilin University of Technology,
Guilin, Guangxi 541004, China
}

\begin{abstract}
Extreme mass-ratio inspirals (EMRIs) are sensitive probes of weak environmental
effects around massive black holes, since such effects can accumulate into
observable gravitational-wave phase shifts. In this work, we study the phase
degeneracy between dark matter halos and accretion disks in eccentric EMRI
waveforms. We model the dark matter (DM) environment with NFW and Beta halo
profiles, and describe the disk using a thin $\alpha$-disk model. Their
distinguishability is quantified through eccentricity-dependent phase
diagnostics and the residual signal-to-noise ratio (SNR) in the
LISA band. Our results show that DM-induced dephasing depends only weakly on
the initial eccentricity $e_0$, whereas disk-induced dephasing is strongly
suppressed as $e_0$ increases. The distinguishability time exhibits a smooth and weakly
non-monotonic dependence on the initial eccentricity, while being more
strongly controlled by the environmental strength.
For the benchmark systems considered here, the DM--disk waveform difference
can be detectable by LISA, and $e_0$ can serve as an auxiliary diagnostic in
addition to the observation duration.
\end{abstract}

\maketitle

\section{Introduction}

The first detections of gravitational waves from compact binary coalescences by ground-based observatories led by the LIGO collaboration have opened a new era of gravitational-wave astronomy\cite{LIGOScientific:2016aoc,LIGOScientific:2016vlm,LIGOScientific:2016emj,LIGOScientific:2018mvr,LIGOScientific:2020ibl,LIGOScientific:2021usb}. In the future, space-based detectors such as the Laser Interferometer Space Antenna (LISA)\cite{Karnesis:2022vdp,LISA:2017pwj,LISA:2022yao}, TianQin\cite{TianQin:2015yph,Liu:2020eko}, and Taiji\cite{Hu:2017mde,Gong:2021gvw} are expected to probe gravitational-wave signals in the millihertz frequency band\cite{Baibhav:2019rsa}. Among the most promising sources for these missions are extreme mass ratio inspirals (EMRIs), whose signals are produced by stellar-mass compact objects orbiting massive or supermassive black holes\cite{Amaro-Seoane:2007osp,Berry:2019wgg,Seoane:2021kkk,Laghi:2021pqk,McGee:2018qwb,Barausse:2014tra,Cardoso:2022whc,Zi:2025lio}. Owing to their long inspiral duration and the large number of accumulated orbital cycles, EMRIs are highly sensitive to small perturbations in the orbital evolution and can therefore carry rich information about the spacetime geometry and astrophysical environments around the central black hole\cite{Yang:2024lmj,Zi:2024jla,Battista:2021rlh}.

A number of studies have investigated EMRIs as probes of astrophysical environments, focusing in particular on how dark matter (DM) distributions and accretion disks modify the orbital evolution and the resulting gravitational waveforms\cite{Yunes2011,Kocsis2011,Eda2015,Li2022,Rahman2024,Speeney2022}. An important issue in this context is that different environmental effects may induce similar phase corrections, leading to potential degeneracies in waveform interpretation\cite{CardosoMaselli2020,Cole2023,BeckerSagunski2023,RiveraReyes2024}. For instance, it has been shown that the waveform signatures produced by different DM halo models can be nearly indistinguishable over short observation times, and that a sufficiently long accumulation of phase difference is required to break this degeneracy\cite{LiQiaoTao2026,Gliorio2025}. Since accretion disks also act as persistent environmental perturbations to the orbital evolution\cite{BarausseRezzolla2008,Derdzinski2019,Derdzinski2021,Basu2024,Duque2025,Liu:2026dug}, they may produce cumulative phase shifts that can be partially degenerate with those induced by DM environments\cite{Cole2023,BeckerSagunski2023}. It is therefore natural to ask whether the phase shifts induced by DM and accretion disks can be distinguished, and how long an observation is required for such a distinction to become possible.

In this work, we address the DM--disk phase degeneracy by combining
phase-level diagnostics with a detector-level residual signal-to-noise ratio (SNR) analysis. We first
introduce two diagnostics based on the initial orbital eccentricity $e_0$. The
first is an eccentricity-sensitivity criterion, which characterizes how
strongly the environmental dephasing responds to changes in $e_0$. This is
motivated by the fact that eccentric motion changes the radial range sampled
by the compact object, the time spent near different orbital radii, and its
relative velocity with respect to the surrounding matter. Accordingly, dark
matter halos and accretion disks, which affect the orbital evolution through
different physical mechanisms, are not expected to induce the same
eccentricity dependence in the waveform phase.

The second diagnostic is an earliest distinguishability-time criterion. We
define the distinguishability time $T_{\rm dis}$ as the earliest observation
time at which the accumulated phase difference between the dark-matter and
disk-induced waveforms reaches a fiducial threshold of $1~\mathrm{rad}$. The
eccentricity-sensitivity criterion identifies whether the two environments
respond differently to $e_0$, while the distinguishability-time criterion
quantifies how long the signal must be observed before this difference becomes
phase-resolvable. To further assess whether this waveform difference can be
distinguished in a realistic space-based detector, we also compute the
residual signal-to-noise ratio between the dark-matter and accretion-disk
waveforms using the LISA noise power spectral density. This residual-SNR
analysis connects the phase-level distinguishability to detector sensitivity
and allows us to evaluate the detectability of the DM--disk waveform
difference in a noise-weighted sense.
Together, these diagnostics allow us to assess the DM--disk distinguishability
from three complementary aspects: the eccentricity dependence of the
environmental phase response, the observation time required for the
accumulated phase difference to become resolvable, and the residual
detectability of the waveform difference in the LISA band.

The remainder of this paper is organized as follows. The waveform construction
adopted in this work is first described. The incorporation of the DM
halo and accretion disk effects into the waveform model is then presented
through their modifications to the orbital evolution. In the results section,
the eccentricity dependence of the environmental dephasing is compared for the
DM and accretion disk models. The phase distinguishability between the
two environmental effects is then investigated using the distinguishability
time $T_{\rm dis}$, followed by an evaluation of their residual signal-to-noise
ratio based on the LISA noise power spectral density. Finally, the main
conclusions are summarized and the implications of the results are discussed.

\section{Numerical Kludge Waveform}
\label{sec:numerical_kludge}

We generate the gravitational waveform using the Numerical Kludge method \cite{Babak2007,Chua2017}. The basic idea of the Numerical Kludge approach is to first compute the trajectory of the secondary compact object and then map this trajectory to gravitational radiation through an approximate quadrupole formula\cite{Tu:2023xab}. To describe the trajectory of the secondary compact object, we first specify the background spacetime in which it moves. In this work, we adopt a static and spherically symmetric metric of the form
\begin{equation}
ds^2
=
-f(r)dt^2
+
f^{-1}(r)dr^2
+
r^2d\Omega^2 ,
\label{eq:sss_metric}
\end{equation}
where $d\Omega^2=d\theta^2+\sin^2\theta d\phi^2$, and $f(r)$ is the metric function determined by the central black hole and its surrounding environment\cite{Xu:2018wow}.

In this background spacetime, the trajectory of the secondary compact object is governed by the geodesic equation
\begin{equation}
\frac{d u^\alpha}{d\tau}
+
\Gamma^\alpha_{\mu\nu}u^\mu u^\nu
=
0,
\label{eq:geodesic_equation}
\end{equation}
where $u^\alpha=dx^\alpha/d\tau$ denotes the four-velocity of the secondary compact object, and $\tau$ is its proper time. Due to the spherical symmetry of the background spacetime, we restrict the motion to the equatorial plane without loss of generality, namely $\theta=\pi/2$ and $\dot{\theta}=0$.

For the static and spherically symmetric spacetime considered here, the motion
of the secondary compact object admits two conserved quantities associated with
the timelike and axial Killing vectors
\cite{Chandrasekhar1983,MTW1973}.
These quantities are the specific energy $\epsilon$ and the specific angular
momentum $l$, defined as
\begin{subequations}
\begin{align}
u_t
&=
-\frac{E}{\mu}
=
-\epsilon ,
\label{eq:specific_energy}
\\
u_\phi
&=
\frac{L}{\mu}
=
l ,
\label{eq:specific_angular_momentum}
\end{align}
\end{subequations}
where $E$ and $L$ denote the orbital energy and angular momentum, respectively,
and $\mu$ is the mass of the secondary compact object.

For a massive particle, the four-velocity satisfies the normalization condition
\begin{equation}
g_{\mu\nu}u^\mu u^\nu
=
-1 .
\label{eq:four_velocity_normalization}
\end{equation}
Using this condition together with the definitions of $\epsilon$ and $l$, one
obtains the radial equation of motion
\begin{equation}
\left(
\frac{dr}{d\tau}
\right)^2
+
f(r)
\left(
1+\frac{l^2}{r^2}
\right)
=
\epsilon^2 .
\label{eq:radial_equation}
\end{equation}

To describe bound eccentric orbits with relativistic precession, we introduce a
Keplerian-like parametrization in terms of the semi-latus rectum $p$, the
eccentricity $e$, and the radial phase variable $\chi$
\cite{BarackCutler2004,Babak2007}:
\begin{equation}
r(\chi)
=
\frac{p}{1+e\cos\chi}.
\label{eq:r_chi}
\end{equation}
Here $\chi$ parametrizes the radial motion. In this parametrization, $\chi=0$
corresponds to the pericenter, while $\chi=\pi$ corresponds to the apocenter. The two radial turning points are
\begin{equation}
r_{\rm p}
=
\frac{p}{1+e},
\qquad
r_{\rm a}
=
\frac{p}{1-e}.
\label{eq:turning_points}
\end{equation}

At the turning points, the radial velocity vanishes,
\begin{equation}
\left.
\frac{dr}{d\tau}
\right|_{r=r_{\rm p}}
=
\left.
\frac{dr}{d\tau}
\right|_{r=r_{\rm a}}
=
0 .
\label{eq:turning_velocity}
\end{equation}
Substituting this condition into Eq.~\eqref{eq:radial_equation}, we obtain
\begin{subequations}
\begin{align}
\epsilon^2
&=
f(r_{\rm p})
\left(
1+\frac{l^2}{r_{\rm p}^2}
\right),
\label{eq:turning_condition_rp}
\\
\epsilon^2
&=
f(r_{\rm a})
\left(
1+\frac{l^2}{r_{\rm a}^2}
\right).
\label{eq:turning_condition_ra}
\end{align}
\end{subequations}

These two equations determine the conserved quantities $\epsilon$ and $l$ once
the orbital parameters $(p,e)$ and the metric function $f(r)$ are specified.

Solving Eqs.~\eqref{eq:turning_condition_rp} and
\eqref{eq:turning_condition_ra}, we can express the constants of motion
$l^2$ and $\epsilon^2$ in terms of the two radial turning points:
\begin{subequations}
\begin{align}
l^2
&=
\frac{
r_{\rm p}^2 r_{\rm a}^2
\left[
f(r_{\rm a})-f(r_{\rm p})
\right]
}{
f(r_{\rm p})r_{\rm a}^2
-
f(r_{\rm a})r_{\rm p}^2
},
\label{eq:l_turning_points}
\\
\epsilon^2
&=
\frac{
f(r_{\rm p})f(r_{\rm a})
\left(
r_{\rm a}^2-r_{\rm p}^2
\right)
}{
f(r_{\rm p})r_{\rm a}^2
-
f(r_{\rm a})r_{\rm p}^2
}.
\label{eq:epsilon_turning_points}
\end{align}
\end{subequations}

The temporal and azimuthal components of the four-velocity are given by
\begin{equation}
\frac{dt}{d\tau}
=
\frac{\epsilon}{f(r)},
\qquad
\frac{d\phi}{d\tau}
=
\frac{l}{r^2}.
\label{eq:t_phi_tau}
\end{equation}

Combining Eq.~\eqref{eq:t_phi_tau} with the radial equation
\eqref{eq:radial_equation}, and using the chain rule, we obtain
\begin{subequations}
\begin{align}
\frac{d\phi}{d\chi}
&=
\frac{d\phi}{d\tau}
\frac{d\tau}{dr}
\frac{dr}{d\chi}
\nonumber
\\
&=
\frac{l}{r^2}
\left|
\frac{dr}{d\chi}
\right|
\left[
\epsilon^2
-
f(r)
\left(
1+\frac{l^2}{r^2}
\right)
\right]^{-1/2},
\label{eq:dphi_dchi}
\\
\frac{dt}{d\chi}
&=
\frac{dt}{d\tau}
\frac{d\tau}{dr}
\frac{dr}{d\chi}
\nonumber
\\
&=
\frac{\epsilon}{f(r)}
\left|
\frac{dr}{d\chi}
\right|
\left[
\epsilon^2
-
f(r)
\left(
1+\frac{l^2}{r^2}
\right)
\right]^{-1/2}.
\label{eq:dt_dchi}
\end{align}
\end{subequations}

Here $r=r(\chi)$ is given by Eq.~\eqref{eq:r_chi}. The absolute value is used
to ensure that the coordinate time $t$ and the azimuthal phase $\phi$ increase
continuously over a complete radial cycle.

The coordinate-time radial period $T$ and the pericenter precession angle
$\Delta\phi$ accumulated over one radial cycle are then defined as
\begin{subequations}
\begin{align}
T
&=
\int_{0}^{2\pi}
\frac{dt}{d\chi}
d\chi ,
\label{eq:radial_period}
\\
\Delta\phi
&=
\int_{0}^{2\pi}
\frac{d\phi}{d\chi}
d\chi
-
2\pi .
\label{eq:pericenter_precession}
\end{align}
\end{subequations}

After obtaining the orbital motion, we construct the Numerical Kludge waveform
by mapping the curved-spacetime trajectory to an auxiliary flat-space
trajectory with the same spherical coordinates
\cite{Babak2007}.
In the equatorial plane, this trajectory can be written as
\begin{equation}
x(t)
=
r(t)\cos\phi(t),
\qquad
y(t)
=
r(t)\sin\phi(t),
\qquad
z(t)
=
0 .
\label{eq:flat_space_trajectory}
\end{equation}

Here $r(t)$ and $\phi(t)$ are obtained from the relations $dt/d\chi$ and
$d\phi/d\chi$ derived above.

The mass quadrupole moment of the secondary compact object is then given by
\begin{equation}
I^{ij}(t)
=
\mu
\left[
x^i(t)x^j(t)
-
\frac{1}{3}\delta^{ij}r^2(t)
\right],
\label{eq:quadrupole_moment}
\end{equation}
where $\mu$ is the mass of the secondary compact object. In the leading-order
quadrupole approximation, the transverse-traceless metric perturbation is
\cite{Maggiore2008}
\begin{equation}
h_{ij}^{\rm TT}(t)
=
\frac{2}{D}
\frac{d^2 I_{ij}^{\rm TT}(t)}{dt^2},
\label{eq:quadrupole_waveform}
\end{equation}
where $D$ is the luminosity distance to the source. Finally, the two gravitational-wave polarizations are obtained by projecting $h_{ij}^{\rm TT}$ onto the plus and cross polarization tensors\cite{Will:2016sgx,Maselli:2021men,Liang:2022gdk},
\begin{equation}
h_+(t)
=
\frac{1}{2}
e_+^{ij}h_{ij}^{\rm TT}(t),
\qquad
h_\times(t)
=
\frac{1}{2}
e_\times^{ij}h_{ij}^{\rm TT}(t),
\label{eq:polarizations}
\end{equation}
where $e_+^{ij}$ and $e_\times^{ij}$ are the plus and cross polarization tensors defined with respect to the line of sight.

\section{Incorporating Environmental Effects into the Waveform Model}

\subsection{Dark Matter Halo}

In this work, the gravitational effect of the DM halo is described by a
Newtonian-order modification of the black-hole metric function $f(r)$. The difference between this Newtonian-order treatment and the relativistic
dark-matter metric\cite{Zhao:2023tyo} is discussed in
Appendix~\ref{app}.
In this case, the metric function is written as\cite{Xu:2018wow}
\begin{equation}
f(r)
=
f_{\rm DM}(r)
-
\frac{2M}{r},
\label{eq:f_dm_general}
\end{equation}
where $M$ denotes the mass of the central black hole, while $f_{\rm DM}(r)$ encodes the contribution from the surrounding DM halo. In the absence of the DM halo, one has $f_{\rm DM}(r)=1$, and Eq.~\eqref{eq:f_dm_general} reduces to the Schwarzschild metric function $f(r)=1-2M/r$.

The form of $f_{\rm DM}(r)$ is determined by the internal mass distribution of the DM halo. In the Newtonian limit, the tangential velocity of a test particle moving in a spherically symmetric gravitational field is related to the enclosed DM mass by
\begin{equation}
v_{\rm tg}^2(r)
\simeq
\frac{M_{\rm DM}(r)}{r}.
\label{eq:vtg_mass_relation}
\end{equation}
On the other hand, for a static and spherically symmetric metric, the tangential velocity can be expressed in terms of the metric function as\cite{Matos:2000ki}
\begin{equation}
v_{\rm tg}^2(r)
=
\frac{r}{\sqrt{f(r)}}
\frac{d\sqrt{f(r)}}{dr}
=
r
\frac{d\ln\sqrt{f(r)}}{dr}.
\label{eq:vtg_metric_relation}
\end{equation}
Applying this relation to the dark-matter contribution and using Eq.~\eqref{eq:vtg_mass_relation}, we obtain
\begin{equation}
\frac{d\ln\sqrt{f_{\rm DM}(r)}}{dr}
=
\frac{M_{\rm DM}(r)}{r^2}.
\label{eq:f_dm_ode}
\end{equation}
Solving this ordinary differential equation gives
\begin{equation}
f_{\rm DM}(r)
=
\exp
\left[
2
\int
\frac{M_{\rm DM}(r)}{r^2}
dr
\right].
\label{eq:f_dm_integral}
\end{equation}

To obtain explicit metric functions, one needs to specify the DM density profile. In this work, we consider two widely used spherically symmetric halo profiles: the Navarro--Frenk--White (NFW) profile and the Beta profile. Their density distributions are given by\cite{Navarro:1994hi,Navarro:1995iw,Navarro:1996gj,Cavaliere:1976tx}
\begin{subequations}
\begin{align}
\rho_{\rm NFW}(r)
&=
\frac{\rho_0}
{
(r/h)
\left(
1+r/h
\right)^2
},
\label{eq:rho_nfw}
\\
\rho_{\rm Beta}(r)
&=
\frac{\rho_0}
{
\left[
1+
(r/h)^2
\right]^{3/2}
},
\label{eq:rho_beta}
\end{align}
\end{subequations}
where $\rho_0$ denotes the characteristic density and $h$ is the characteristic radius of the halo. For a spherically symmetric distribution, the cumulative DM mass enclosed within radius $r$ is calculated as
\begin{equation}
M_{\rm DM}(r)
=
4\pi
\int_{0}^{r}
\rho(r')\,r'^2\,dr' .
\label{eq:mdm_enclosed}
\end{equation}

Substituting Eqs.~\eqref{eq:rho_nfw} and \eqref{eq:rho_beta} into Eq.~\eqref{eq:mdm_enclosed}, one obtains the corresponding cumulative DM masses. Further substituting these mass functions into Eq.~\eqref{eq:f_dm_integral}, we obtain the effective metric functions for the full gravitational system\cite{Li:2025eln,Qiao:2024ehj,Liu:2023xtb}:
\begin{subequations}
\begin{align}
f_{\rm NFW}(r)
&=
\left(
1+\frac{r}{h}
\right)^{-\frac{8\pi k}{r}}
-
\frac{2M}{r},
\label{eq:f_nfw}
\\
f_{\rm Beta}(r)
&=
\exp
\left[
-\frac{8\pi k}{r}
\operatorname{arcsinh}
\left(
\frac{r}{h}
\right)
\right]
-
\frac{2M}{r},
\label{eq:f_beta}
\end{align}
\end{subequations}
where
\begin{equation}
k
=
\rho_0 h^3
\label{eq:k_definition}
\end{equation}
characterizes the strength of the DM halo contribution. Equations~\eqref{eq:f_nfw} and \eqref{eq:f_beta} provide the explicit analytical forms of the metric function for the NFW and Beta DM halo profiles, respectively.

In addition to modifying the conservative background spacetime, the DM halo also induces dissipative effects on the motion of the secondary compact object. As the small black hole moves through the DM halo, it experiences a gravitational drag force known as dynamical friction. This force arises from the gravitational interaction between the moving compact object and the surrounding DM particles, leading to a loss of orbital energy and angular momentum.

For a collisionless DM medium, the dynamical friction force can be written as\cite{Chandrasekhar:1943ys,Cardoso:2020iji}
\begin{equation}
\mathbf{F}_{\rm DF}
=
-
\frac{
4\pi \mu^2 \rho_{\rm DM}(r)\ln\Lambda
}{
v^3
}
\mathbf{v},
\label{eq:dynamical_friction_force}
\end{equation}
where $\mathbf{v}$ is the velocity of the secondary compact object, $v=|\mathbf{v}|$, $\rho_{\rm DM}(r)$ is the local DM density, and $\ln\Lambda$ is the Coulomb logarithm. In this work, we take $\ln\Lambda\simeq 3$\cite{Eda2015}. For completeness, the effect of adopting a larger value,
$\ln\Lambda\simeq 10$\cite{Yue:2019}, is further investigated in Appendix~\ref{app:lambda10}.

The corresponding instantaneous rates of energy and angular momentum loss induced by dynamical friction are
\begin{subequations}
\begin{align}
\left(
\frac{dE}{dt}
\right)_{\rm DF}
&=
\mathbf{F}_{\rm DF}\cdot\mathbf{v},
\label{eq:dedt_df}
\\
\left(
\frac{dL}{dt}
\right)_{\rm DF}
&=
\left(
\mathbf{r}\times\mathbf{F}_{\rm DF}
\right)_z .
\label{eq:dldt_df}
\end{align}
\end{subequations}
Here the subscript $z$ denotes the component perpendicular to the orbital plane. By substituting the orbital trajectory derived in the previous section and the corresponding density profiles into Eqs.~\eqref{eq:dedt_df} and \eqref{eq:dldt_df}, one obtains the instantaneous energy and angular momentum loss rates associated with the NFW and Beta halo models.

Besides dynamical friction, the secondary compact object, assumed to be a black hole, can also accrete DM particles from its surrounding environment. We model this process using the Bondi--Hoyle--Lyttleton accretion prescription. The accretion rate of the secondary mass is given by\cite{Bondi:1952ni,Macedo:2013qea,Bondi:1944rnk,Edgar:2004mk}
\begin{equation}
\dot{\mu}
=
\frac{
4\pi \rho_{\rm DM}(r)\mu^2
}{
\left(
v^2+c_s^2
\right)^{3/2}
},
\label{eq:bhl_accretion_rate}
\end{equation}
where $c_s=\sqrt{\delta P/\delta\rho}$ is the sound speed of the DM medium. Near a supermassive black hole, the orbital velocity of the secondary compact object is typically much larger than the sound speed, $v\gg c_s$. In this limit, Eq.~\eqref{eq:bhl_accretion_rate} can be approximated as
\begin{equation}
\dot{\mu}
\simeq
\frac{
4\pi \rho_{\rm DM}(r)\mu^2
}{
v^3
}.
\label{eq:bhl_accretion_rate_approx}
\end{equation}

In this work, we assume isotropic accretion from a non-rotating DM medium. Under this assumption, the accretion process does not transfer angular momentum to the secondary compact object, so that
\begin{equation}
\left(
\frac{dL}{dt}
\right)_{\rm ACC}
=
0 .
\label{eq:dldt_acc}
\end{equation}
Moreover, according to adiabatic invariance, the orbital eccentricity remains unchanged during the accretion process\cite{Hughes:2018qxz,Blachier:2023ygh},
\begin{equation}
\left(
\frac{de}{dt}
\right)_{\rm ACC}
=
0 .
\label{eq:dedt_acc_eccentricity}
\end{equation}
The corresponding variation of the orbital energy caused by the mass growth of the secondary compact object is given by
\begin{equation}
\left(
\frac{dE}{dt}
\right)_{\rm ACC}
=
\frac{\dot{\mu}}{\mu}E
+
\mu\dot{\epsilon}
=
\frac{\dot{\mu}}{\mu}
\left(
E
-
l
\frac{dp}{dl}
\frac{dE}{dp}
\right).
\label{eq:dedt_acc}
\end{equation}

The total evolution of the EMRI system is then driven by the combined effects
of gravitational-wave radiation reaction, dynamical friction, and DM
accretion. The total rates of change of the orbital energy and angular
momentum are written as
\begin{subequations}
\begin{align}
\left\langle
\frac{dE}{dt}
\right\rangle_{\rm total}
&=
\left\langle
\frac{dE}{dt}
\right\rangle_{\rm GW}
+
\left\langle
\frac{dE}{dt}
\right\rangle_{\rm DF}
+
\left\langle
\frac{dE}{dt}
\right\rangle_{\rm ACC},
\label{eq:dedt_total}
\\
\left\langle
\frac{dL}{dt}
\right\rangle_{\rm total}
&=
\left\langle
\frac{dL}{dt}
\right\rangle_{\rm GW}
+
\left\langle
\frac{dL}{dt}
\right\rangle_{\rm DF}.
\label{eq:dldt_total}
\end{align}
\end{subequations}

Here $\langle\cdots\rangle$ denotes the time average over one radial
period $T$. Explicitly, for an instantaneous quantity
$X$, the orbital average is evaluated as
\begin{equation}
\left\langle X \right\rangle
\equiv
\frac{1}{T}\int_{0}^{T}X(t)\,dt
=
\frac{
\displaystyle
\int_{0}^{2\pi}
X(\chi)\frac{dt}{d\chi}\,d\chi
}{
\displaystyle
\int_{0}^{2\pi}
\frac{dt}{d\chi}\,d\chi
},
\qquad
T=
\int_{0}^{2\pi}
\frac{dt}{d\chi}\,d\chi .
\label{eq:orbit_average}
\end{equation}

Here,
$\chi$ denotes the relativistic anomaly, and
$dt/d\chi$ is determined by the relativistic orbital equations, in
particular Eq.~\eqref{eq:dt_dchi}.
In the Newtonian limit, Eq.~\eqref{eq:orbit_average}
reduces to the orbital-averaging prescription used in Ref.~\cite{Yue:2019}
. The gravitational-wave terms in Eqs.~\eqref{eq:dedt_total} and \eqref{eq:dldt_total}
 are adopted directly as the leading-order orbit-averaged fluxes,
corresponding to Eqs.~(11) and (12) of Ref.~\cite{Yue:2019}, while the dynamical-friction terms are averaged following the
same time-averaging principle as Eqs.~(15) and (16) of that reference. The
dark-matter accretion contribution is evaluated using the same orbital-average
definition in Eq.~\eqref{eq:orbit_average}.

\subsection{Accretion Disk}

We next consider the environmental effect induced by an accretion disk surrounding the supermassive black hole. In this work, we assume a geometrically thin $\alpha$ accretion disk and take the mass accretion rate onto the central black hole to be approximately constant over the timescale of interest. The accretion rate satisfies\cite{acc_book}
\begin{equation}
\dot{M}
\simeq
3\pi\nu\Sigma ,
\label{eq:disk_accretion_rate}
\end{equation}
where $\nu$ is the kinematic viscosity of the disk and $\Sigma$ is the disk surface density. The mass accretion rate
$\dot{M}$
is used only to characterize the disk structure. The disk
self-gravity and the variation of the central black-hole mass are neglected,
and the disk effect is included only through migration and eccentricity
damping.

In the $\alpha$-disk model, the kinematic viscosity is commonly parameterized as
\begin{equation}
\nu
=
\alpha_{\rm disk} c_s H(r),
\label{eq:alpha_viscosity}
\end{equation}
where $\alpha_{\rm disk}$ is a dimensionless viscosity parameter, typically in the range $\alpha_{\rm disk}\in[0.001,0.1]$\cite{Jiang:2019bxn}, $c_s=H\Omega_{\rm K}=hr\Omega_{\rm K}$ is the isothermal sound speed,
$h=H/r$ is the disk aspect ratio, $H(r)$ is the disk scale height, and
\begin{equation}
\Omega_{\rm K}
=
\sqrt{\frac{M}{r^3}}
\label{eq:keplerian_frequency}
\end{equation}
is the Keplerian angular frequency.

For a standard $\alpha$ disk, the surface density and the disk aspect ratio can be written as
\cite{ShakuraSunyaev1973,FrankKingRaine2002,Liu2026EMRIAGN}
\begin{equation}
\Sigma_{\alpha}(r)
=
5.4\times10^2
\left(
\frac{0.1}{\alpha_{\rm disk}}
\right)
\left(
\frac{0.1}{f_{\rm Edd}}
\right)
\left(
\frac{\epsilon}{0.1}
\right)
\left(
\frac{r}{10M}
\right)^{3/2}
\left(
1-\sqrt{\frac{r_{\rm in}}{r}}
\right)^{-1}
\ {\rm g\,cm^{-2}},
\label{eq:sigma_alpha_disk}
\end{equation}
and
\begin{equation}
h_{\alpha}(r)
=
0.15
\left(
\frac{f_{\rm Edd}}{0.1}
\right)
\left(
\frac{0.1}{\epsilon}
\right)
\left(
\frac{r}{10M}
\right)^{-1}
\left(
1-\sqrt{\frac{r_{\rm in}}{r}}
\right),
\label{eq:h_alpha_disk}
\end{equation}
where $f_{\rm Edd}$ is the Eddington ratio, $\epsilon$ is the radiative efficiency, and $r_{\rm in}$ denotes the inner edge of the accretion disk.

The structural parameters of the $\alpha$ disk can be approximated by power-law profiles
\cite{FrankKingRaine2002,Kocsis2011,Liu2026EMRIAGN}.
Since this work focuses on the orbital evolution of the EMRI during the late inspiral stage,
when the secondary compact object mainly moves in the inner region of the disk,
we retain the correction terms associated with the inner boundary condition
\cite{FrankKingRaine2002,Liu2026EMRIAGN}.
The surface density and aspect ratio are therefore parameterized as
\begin{equation}
\Sigma(r)
=
\Sigma_0
\left(
\frac{r}{10M}
\right)^{-\Sigma_p}
\left(
1-\sqrt{\frac{r_{\rm in}}{r}}
\right)^{-1}
\ {\rm g\,cm^{-2}},
\label{eq:sigma_parametrized}
\end{equation}
and
\begin{equation}
h(r)
=
h_0
\left(
\frac{r}{10M}
\right)^{(2\Sigma_p-1)/4}
\left(
1-\sqrt{\frac{r_{\rm in}}{r}}
\right).
\label{eq:h_parametrized}
\end{equation}
Here $\Sigma_0$ and $h_0$ are representative inner-disk values inferred from astrophysical observations, with $\Sigma_0\in[10^3,10^6]$ and $h_0\in[0.01,0.1]$. For an $\alpha$ disk, we take $\Sigma_p=-3/2$. In this work, the inner radius $r_{\rm in}$ is identified with the radius of the innermost stable circular orbit.

Before constructing the effective torque or drag exerted by the accretion disk on the secondary compact object, one needs to compare the relative velocity between the secondary compact object and the disk gas with the local sound speed. The form of the environmental interaction depends on whether the motion is subsonic or supersonic. In general, the relative velocity receives contributions from both the orbital eccentricity and the inclination of the orbit with respect to the disk mid-plane. We adopt the approximation\cite{Liu2026EMRIAGN}
\begin{equation}
v_{\rm rel}
\sim
v_{\rm K}
\sqrt{
e^2+\sin^2\iota_{\rm sd}
}
\sim
r\Omega_{\rm K}
\sqrt{
e^2+\sin^2\iota_{\rm sd}
},
\label{eq:relative_velocity_disk}
\end{equation}
where $\iota_{\rm sd}$ is the angle between the orbital plane of the secondary compact object and the disk mid-plane, and $v_{\rm K}=r\Omega_{\rm K}$ is the Keplerian velocity.

We assume that once the EMRI enters the LISA band, corresponding to the inner region of the accretion disk, the secondary compact object is fully embedded in the disk and satisfies $\iota_{\rm sd}<h$. Since the inclination damps rapidly and the orbit circularizes efficiently in wet EMRIs, both $\iota_{\rm sd}$ and $e$ are typically small. The Mach number is then given by\cite{Pan:2021ksp,Basu2024}
\begin{equation}
\mathcal{M}
\equiv
\frac{v_{\rm rel}}{c_s}
\simeq
\frac{
r\Omega_{\rm K}
\sqrt{
e^2+\sin^2\iota_{\rm sd}
}
}{
h r\Omega_{\rm K}
}
=
\frac{
\sqrt{
e^2+\sin^2\iota_{\rm sd}
}
}{h}.
\label{eq:mach_number_disk}
\end{equation}

Under typical conditions, the migration regime of the secondary compact object is determined by the Mach number $\mathcal{M}$. When $\mathcal{M}>1$, the migration occurs in the supersonic regime, whereas when $\mathcal{M}<1$, it occurs in the subsonic regime. Since the value of $\mathcal{M}$ depends on the specific disk and orbital parameters, we adopt a prescription based on dynamical friction theory that is applicable to both regimes. In the small-inclination limit, $\iota_{\rm sd}<h$, the environmental impact on the orbital evolution can be characterized by the following inverse timescales\cite{2020MNRAS}:
\begin{subequations}
\begin{align}
\tau_e^{-1}
&\equiv
-\frac{\dot{e}}{e}
=
0.780\,t_{\rm wave}^{-1}
\left[
1+
\frac{1}{15}
\left(
\left(\frac{e}{h}\right)^2
+
\left(\frac{\iota_{\rm sd}}{h}\right)^2
\right)^{3/2}
\right]^{-1},
\label{eq:tau_e_disk}
\\
\tau_i^{-1}
&\equiv
-\frac{\dot{\iota}_{\rm sd}}{\iota_{\rm sd}}
=
0.544\,t_{\rm wave}^{-1}
\left[
1+
\frac{1}{21.5}
\left(
\left(\frac{e}{h}\right)^2
+
\left(\frac{\iota_{\rm sd}}{h}\right)^2
\right)^{3/2}
\right]^{-1},
\label{eq:tau_i_disk}
\\
\tau_a^{-1}
&\equiv
-\frac{\dot{a}}{a}
=
4.25h^2\,t_{\rm wave}^{-1}
\left[
1+
\frac{4.25}{21}
\left(
\left(\frac{e}{h}\right)^2
+
\left(\frac{\iota_{\rm sd}}{h}\right)^2
\right)^{1/2}
\right]^{-1}.
\label{eq:tau_a_disk}
\end{align}
\end{subequations}
Here $t_{\rm wave}^{-1}$ is the inverse characteristic migration timescale, defined as
\begin{equation}
t_{\rm wave}^{-1}
=
\frac{\mu}{M}
\left(
\frac{\Sigma r^2}{M}
\right)
h^{-4}
\Omega_{\rm K}.
\label{eq:twave_disk}
\end{equation}
The inverse timescale $t_{\rm wave}^{-1}$ controls the strength of the disk-induced migration and damping.

We further clarify how the accretion-disk effect is incorporated into the
eccentric EMRI waveform. Unlike the DM halo case, the accretion disk does not
modify the background spacetime in this work. Instead, its effect is included
through the secular evolution of the orbital parameters. The disk-induced
migration and eccentricity damping rates are given by
Eqs.%
\hyperref[eq:tau_a_disk]{(\ref*{eq:tau_a_disk})}
and
\hyperref[eq:tau_e_disk]{(\ref*{eq:tau_e_disk})},
which lead to\cite{2020MNRAS}
\begin{equation}
\left(\frac{da}{dt}\right)_{\rm disk}
=
-a\tau_a^{-1},
\qquad
\left(\frac{de}{dt}\right)_{\rm disk}
=
-e\tau_e^{-1}.
\end{equation}

For eccentric EMRIs, the orbital evolution is described by the semi-latus
rectum $p$ and eccentricity $e$. Using
$p=a(1-e^2)$,
the disk contribution is converted into
\begin{equation}
\left(\frac{dp}{dt}\right)_{\rm disk}
=
(1-e^2)
\left(\frac{da}{dt}\right)_{\rm disk}
-
2ae
\left(\frac{de}{dt}\right)_{\rm disk}.
\end{equation}

The total orbital evolution is obtained by adding the disk contribution to
the gravitational-wave radiation reaction,
\begin{equation}
\frac{dp}{dt}
=
\left(\frac{dp}{dt}\right)_{\rm GW}
+
\left(\frac{dp}{dt}\right)_{\rm disk},
\qquad
\frac{de}{dt}
=
\left(\frac{de}{dt}\right)_{\rm GW}
+
\left(\frac{de}{dt}\right)_{\rm disk}.
\end{equation}

The corresponding changes in the orbital energy and angular momentum are
obtained through
\begin{equation}
\left(\frac{dE}{dt}\right)_{\rm disk}
=
\frac{\partial E}{\partial p}
\left(\frac{dp}{dt}\right)_{\rm disk}
+
\frac{\partial E}{\partial e}
\left(\frac{de}{dt}\right)_{\rm disk},
\end{equation}
with an analogous relation for the angular momentum. The disk-induced
migration and damping rates are secular quantities prescribed by the
accretion-disk model and are directly included in the orbital evolution. No additional integration over the relativistic anomaly $\chi$
is applied to the disk contribution.

In contrast, environmental effects described by instantaneous forces or
energy/angular-momentum losses along eccentric trajectories, such as the
DM-induced dynamical friction, require the explicit time-averaging procedure
defined in Eq.%
\eqref{eq:orbit_average}
. During the inspiral, the waveform is generated from a sequence of eccentric
geodesics in the unchanged background spacetime, and the resulting trajectory
is mapped into gravitational radiation using the same Numerical Kludge method
described in Sec.%
\hyperref[sec:numerical_kludge]{\ref*{sec:numerical_kludge}}.
Accordingly, the accretion-disk effect enters the waveform through
\begin{equation}
\{\Sigma(r),h(r)\}
\rightarrow
\{\tau_a,\tau_e\}
\rightarrow
\{p(t),e(t)\}
\rightarrow
\{h_+(t),h_\times(t)\}.
\end{equation}

\section{Results and Discussion}

\begin{table}[htbp]
\centering
\caption{Benchmark parameter choices for the waveform comparisons.}
\label{tab:env_parameters}
\renewcommand{\arraystretch}{1.15}
\begin{tabular*}{\linewidth}{@{\extracolsep{\fill}} l c l}
\hline
Parameter & Symbol & Value \\
\hline
\multicolumn{3}{l}{\textit{General waveform parameters}} \\
\hline
Central black-hole mass & $M$ & $10^6M_{\odot}$ \\
Secondary compact-object mass & $m$ & $10M_{\odot}$ \\
Reduced mass & $\mu$ & $\mu\simeq m\simeq 10M_{\odot}$ \\
Initial semi-latus rectum & $p_0/M$ & $16$ \\
Initial orbital eccentricity & $e_0$ & $[0,\ 0.1]$ \\
Luminosity distance & $D_L$ & $2~{\rm Gpc}$ \\
Line-of-sight inclination angle & $\iota$ & $\pi/4$ \\
Initial pericenter phase & $\zeta$ & $\pi/4$ \\
\hline
\multicolumn{3}{l}{\textit{Dark matter halo parameters}} \\
\hline
Characteristic radius & $h/M$ & $10^7$ \\
Halo strength parameter & $k/M$ & $1000,\ 5000,\ 10000$ \\
\hline
\multicolumn{3}{l}{\textit{Accretion disk parameters}} \\
\hline
Disk--orbit inclination angle & $\iota_{\rm sd}$ & $0$ \\
Aspect-ratio normalization & $h_0$ & $0.05$ \\
Surface-density normalization & $\Sigma_0$ & $10^3,\ 10^4,\ 10^5~{\rm g\,cm^{-2}}$ \\
\hline
\end{tabular*}
\end{table}

For the subsequent waveform calculations, we fix the central black-hole mass to $M=10^6M_{\odot}$ and the secondary compact-object mass to $m=10M_{\odot}$ \cite{AmaroSeoane2007,Babak2017}, with the reduced mass approximated as $\mu\simeq m$ in the extreme-mass-ratio limit. The initial semi-latus rectum is fixed to $p_0/M=16$ \cite{Katz2021,Chua2021}. The luminosity distance is set to $D_L=2~{\rm Gpc}$ \citep{WenGair2005,Babak2017}. Unless otherwise stated, the line-of-sight inclination angle $\iota$ and the initial pericenter phase $\zeta$ are fixed to $\pi/4$ \cite{Katz2021}. To investigate the influence of orbital eccentricity on the waveform evolution, we vary the initial eccentricity in the range $e_0\in[0,0.1]$ \cite{Katz2021,Chua2021}.

To make the parameter choices more transparent, we summarize the benchmark configurations used in the subsequent waveform comparisons in Table~\ref{tab:env_parameters}.

\subsection{DM--Disk Eccentricity Sensitivity Comparison}

As an initial visual comparison, we fix the environmental parameters to $k/M=5000$ for the DM halo and $\Sigma_0=10^4~{\rm g\,cm^{-2}}$ for the accretion disk, and consider three representative initial eccentricities, $e_0\in\{0,\ 0.05,\ 0.1\}$. For each value of $e_0$, we evolve the system for one year and compare the gravitational waveforms generated in the DM halo and accretion disk environments with the corresponding vacuum waveform.

\begin{figure}[htbp]
\centering
\includegraphics[width=0.9\linewidth]{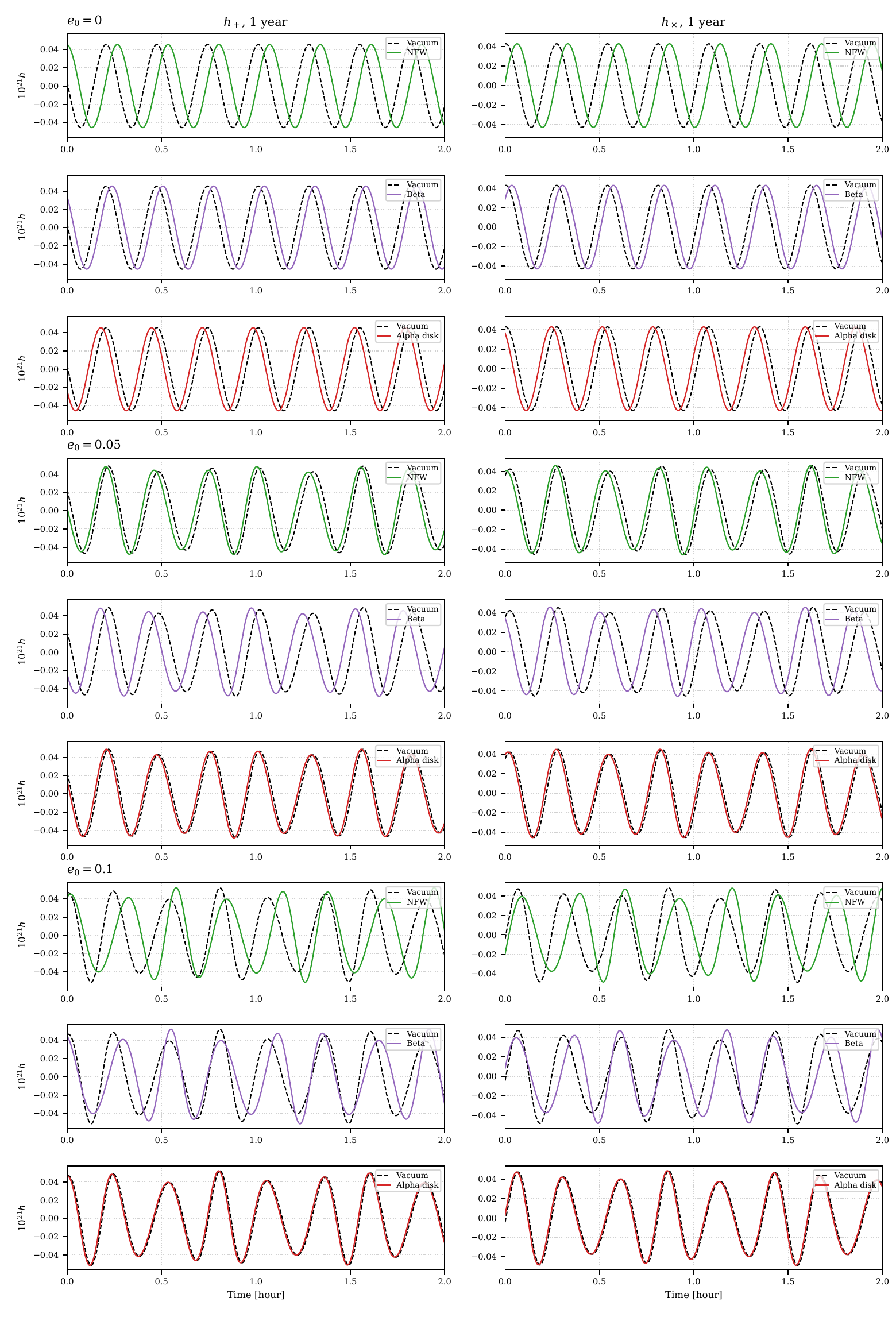}
\caption{Final 2-hour waveform segments extracted after one year of orbital
evolution for $e_0\in\{0,\ 0.05,\ 0.1\}$, with fixed DM parameters $h/M=10^7$, $k/M=5000$, and fixed disk parameters $h_0=0.05$, $\Sigma_0=10^4~{\rm g\,cm^{-2}}$.}
\label{fig:waveform_comparison}
\end{figure}

As shown in Fig.~\ref{fig:waveform_comparison}, we first evolve the EMRI system for one year and then extract the final
2-hour waveform segment as a snapshot for comparison. This short-time waveform
window is used to visualize the accumulated phase differences induced by
different environmental effects after the long-term orbital evolution.
For the accretion disk model, the accumulated waveform phase after one year
of orbital evolution shows a clear dependence on the initial eccentricity.
When $e_0=0.05$ and $e_0=0.1$, the phase difference between the
$\alpha$-disk waveform and the corresponding vacuum waveform is relatively small. In contrast, for the circular case with $e_0=0$, a visible phase dephasing appears. Under the same benchmark parameter configuration, this phase deviation is comparable to that produced by the DM model. For the DM model, the phase evolution varies more mildly as $e_0$ changes, indicating a weaker eccentricity dependence in terms of phase dephasing. However, the DM environment can still modify the waveform amplitude. Since space-based gravitational-wave observations are generally more sensitive to the accumulated phase evolution than to moderate amplitude variations, the amplitude modification is not the main focus of the present analysis.

To quantify how the accumulated environmental dephasing responds to the initial orbital eccentricity, we define the normalized eccentricity response as
\begin{equation}
S_{\rm env}(e_0)
=
\left[
\frac{
|\Delta\Phi_{\rm env}(e_0)|
}{
|\Delta\Phi_{\rm env}(0)|
}
-1
\right]
\times 100\% ,
\label{eq:eccentricity_sensitivity}
\end{equation}
where
\begin{equation}
\Delta\Phi_{\rm env}(e_0)
=
\Phi_{\rm env}(1\,{\rm yr};e_0)
-
\Phi_{\rm vac}(1\,{\rm yr};e_0)
\label{eq:env_dephasing}
\end{equation}
is the accumulated phase difference between the environmental waveform and the corresponding vacuum waveform after one year of evolution. Thus, $S_{\rm env}>0$ indicates that the environmental dephasing is enhanced relative to the circular case, while $S_{\rm env}<0$ indicates that it is suppressed.

\begin{figure}[htbp]
    \centering
    \includegraphics[width=0.86\linewidth]{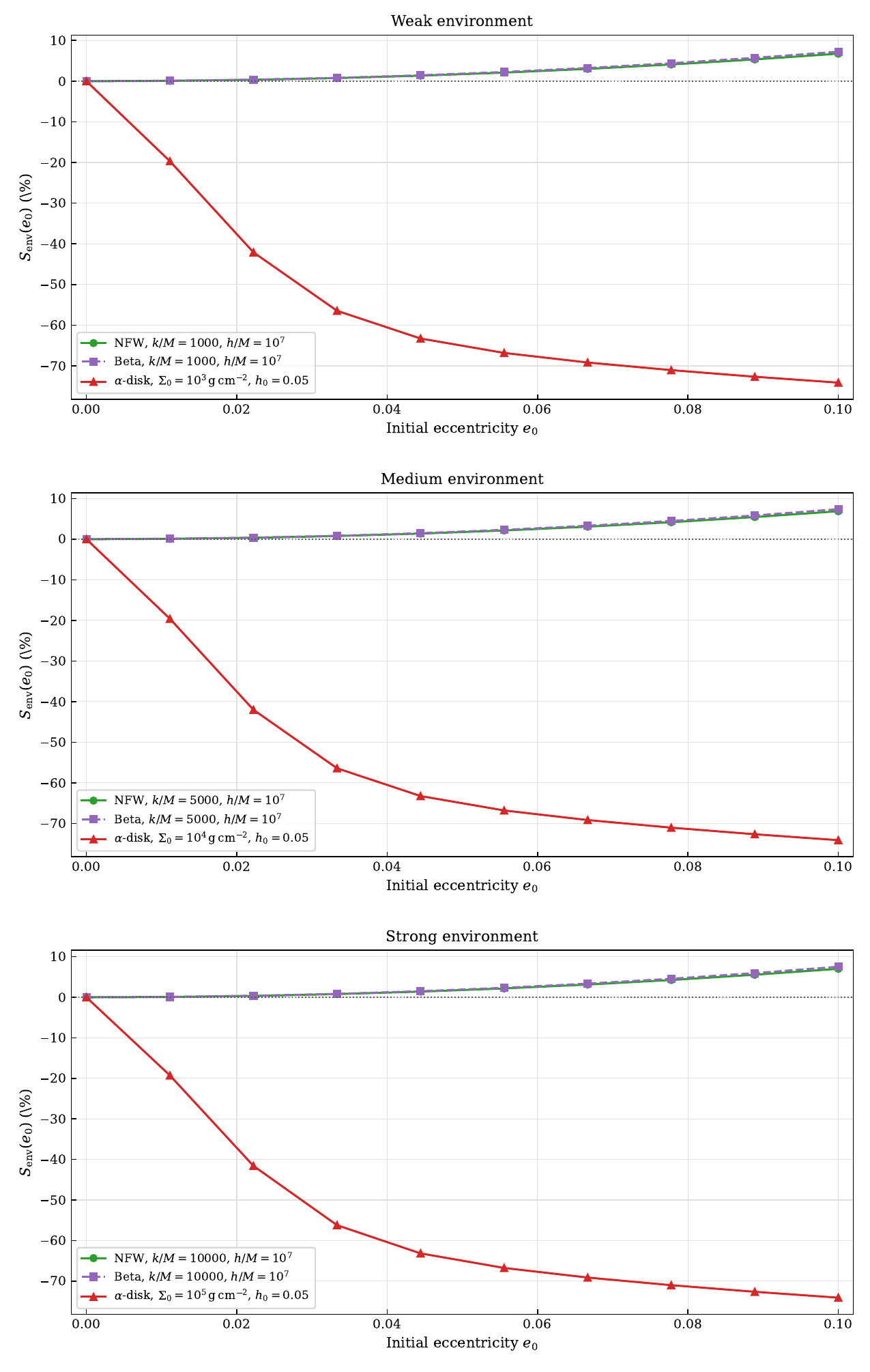}
    \caption{Normalized eccentricity response $S_{\rm env}(e_0)$ of the 1 year environmental dephasing for the NFW, Beta, and $\alpha$-disk models under weak, medium, and strong environmental configurations.}
    \label{fig:eccentricity_sensitivity}
\end{figure}

Using this measure, we compare the eccentricity responses of the DM halo and accretion disk models. The initial eccentricity is sampled at ten values in the range $e_0\in[0,0.1]$. To examine whether the eccentricity dependence is affected by the overall environmental strength, we consider three representative configurations: weak, medium, and strong. For the DM models, these correspond to $k/M=1000$, $5000$, and $10000$ with fixed $h/M=10^7$. For the accretion disk model, they correspond to $\Sigma_0=10^3$, $10^4$, and $10^5~{\rm g\,cm^{-2}}$ with fixed $h_0=0.05$. The resulting dependence of $S_{\rm env}$ on $e_0$ is presented in Fig.~\ref{fig:eccentricity_sensitivity}.

The figure reveals clearly different responses to the initial eccentricity for the DM and accretion disk models. For both the NFW and Beta halo models, $S_{\rm env}(e_0)$ remains close to zero over the whole range $e_0\in[0,0.1]$. Although the Beta model shows a slightly increasing positive response at larger eccentricities, the overall variation remains mild. This indicates that the dark-matter-induced dephasing is only weakly modulated by the initial orbital eccentricity.

In contrast, the $\alpha$-disk model shows a much stronger eccentricity dependence. As $e_0$ increases, $S_{\rm env}(e_0)$ decreases rapidly and reaches a large negative value at $e_0=0.1$, implying that the disk-induced dephasing is significantly suppressed relative to the circular case. This behavior is consistent with the waveform-level trend shown in Fig.~\ref{fig:waveform_comparison}, where the phase deviation becomes weaker for larger $e_0$. The trend appears consistently in the weak, medium, and strong environmental configurations, suggesting that it is a robust feature rather than a result of a particular choice of disk strength. This difference indicates that the eccentricity response provides a useful diagnostic for distinguishing accretion disk effects from DM halo effects: the former is strongly sensitive to $e_0$, whereas the latter remains comparatively insensitive.

\subsection{Eccentricity Dependence of DM--Disk Waveform Distinguishability}

\subsubsection{Phase-Based Distinguishability}

We further investigate the phase distinguishability between the DM and accretion disk environmental effects. Since waveform phase differences accumulate over time, a sufficiently long observation is expected to make the phase discrepancy between the two environmental models distinguishable. Rather than only comparing the final dephasing after a fixed observation time, we focus on how long the signal needs to be observed before the phase difference between the DM and accretion disk models becomes resolvable under different initial eccentricities.

For a given initial eccentricity $e_0$, we define the accumulated phase difference between the two environmental models as
\begin{equation}
\Delta\Phi_{\rm DM-disk}(t;e_0)
=
\left|
\Phi_{\rm DM}(t;e_0)
-
\Phi_{\rm disk}(t;e_0)
\right| ,
\label{eq:dm_disk_phase_difference}
\end{equation}
where $\Phi_{\rm DM}(t;e_0)$ and $\Phi_{\rm disk}(t;e_0)$ denote the gravitational-wave phases generated in the DM halo and accretion disk environments, respectively.

\begin{figure}[htbp]
\centering
\includegraphics[width=0.86\linewidth]{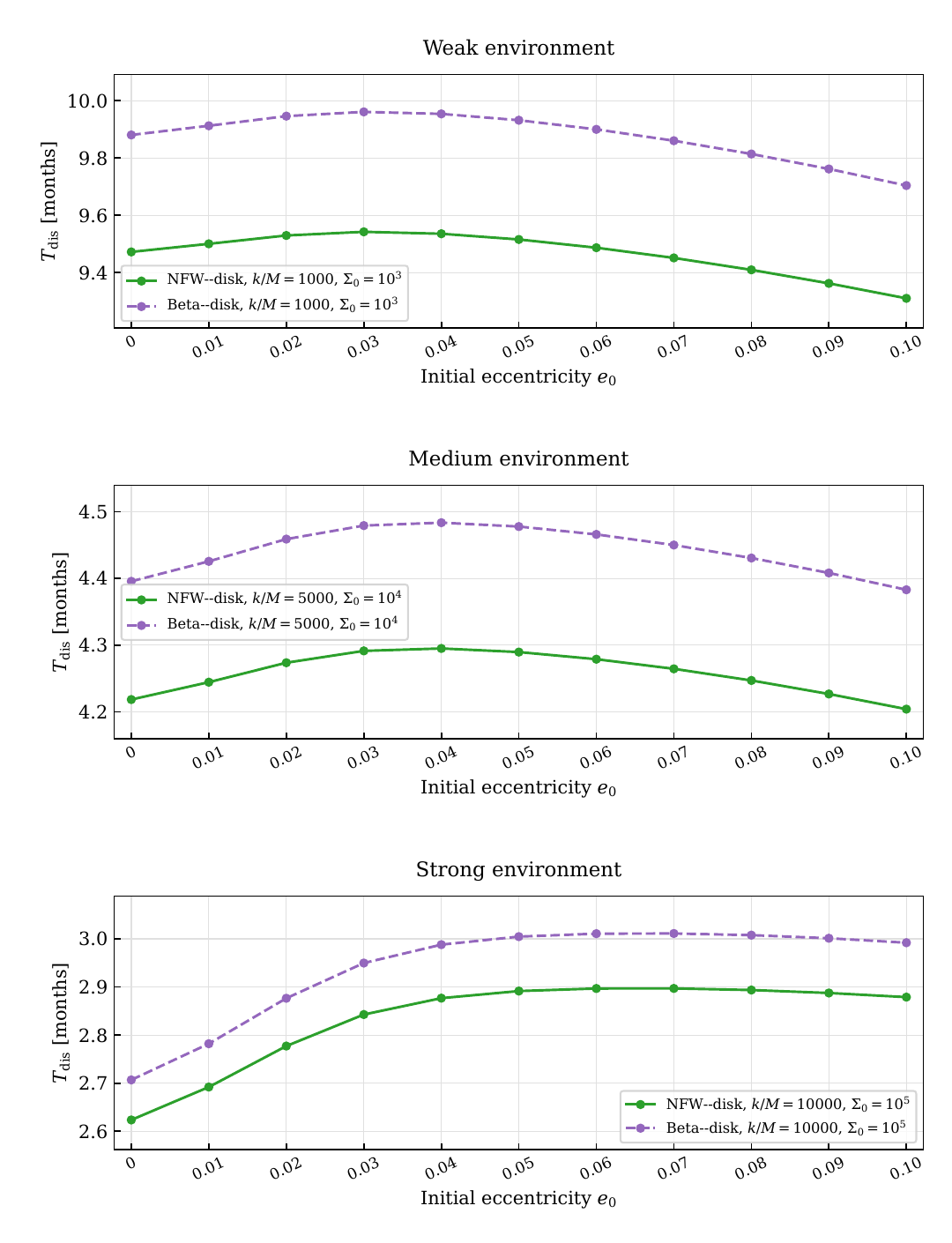}
\caption{Observation time $T_{\rm dis}$ required for the accumulated phase difference between the DM and accretion disk models to reach $1~{\rm rad}$ as a function of the initial eccentricity $e_0$. The three panels correspond to weak, medium, and strong environmental configurations.}
\label{fig:dm_disk_distinguishability_time}
\end{figure}

In this work, we adopt a phase distinguishability threshold of $1~{\rm rad}$. That is, the two environmental models are considered distinguishable in phase once
\begin{equation}
\Delta\Phi_{\rm DM-disk}(t;e_0)
\geq
1~{\rm rad}.
\label{eq:phase_distinguishability_threshold}
\end{equation}
Accordingly, for each value of $e_0$, we define the distinguishability time $T_{\rm dis}(e_0)$ as the earliest observation time at which the accumulated phase difference reaches this threshold:
\begin{equation}
T_{\rm dis}(e_0)
=
\min
\left\{
t:
\Delta\Phi_{\rm DM-disk}(t;e_0)
\geq
1~{\rm rad}
\right\}.
\label{eq:distinguishability_time}
\end{equation}
This quantity allows us to evaluate how the initial eccentricity affects the observation time required to distinguish the phase evolution induced by DM from that induced by the accretion disk. Following the parameter settings introduced above, we classify the environmental effects into weak, medium, and strong configurations. We then calculate the distinguishability time $T_{\rm dis}$ for each initial eccentricity and plot its dependence on $e_0$, as shown in Fig.~\ref{fig:dm_disk_distinguishability_time}.

For all three environmental configurations, the distinguishability time
$T_{\rm dis}$
exhibits a smooth and weakly non-monotonic dependence on the initial
eccentricity. In the weak and medium environments,
$T_{\rm dis}$
slightly increases at small eccentricities, reaches a broad maximum around
$e_0\simeq0.03$--$0.04$,
and then gradually decreases. For the strong environment, the maximum shifts
to
$e_0\simeq0.06$--$0.07$,
followed by a mild decrease at larger eccentricities. This indicates that a
small nonzero eccentricity does not significantly reduce the distinguishability
time, and the eccentricity dependence remains continuous over the considered
range.

The NFW--disk and Beta--disk cases show similar eccentricity-dependent trends,
while stronger environmental effects generally lead to shorter
distinguishability times. These results suggest that the environmental
strength plays a dominant role in determining the phase distinguishability,
whereas the initial eccentricity provides only a moderate modulation.

\subsubsection{LISA Residual SNR for DM--Disk Distinguishability}

To simulate the distinguishability of the two environmental effects in a
realistic future detector, we take LISA as a representative example and
evaluate their detectability under the LISA noise power spectral density.
The residual signal-to-noise ratio is adopted as the quantitative measure
of the DM--disk waveform distinguishability, following the standard
noise-weighted inner-product formalism used in gravitational-wave data
analysis~\cite{CutlerFlanagan1994,JaranowskiKrolak2012,LindblomOwenBrown2008}.
For a given initial eccentricity $e_0$ and observation time $T_{\rm obs}$,
the residual waveforms are defined as
\begin{equation}
    \Delta h_{+}(t;e_0,T_{\rm obs})
    =
    h_{+,{\rm DM}}(t;e_0,T_{\rm obs})
    -
    h_{+,{\rm disk}}(t;e_0,T_{\rm obs}),
\end{equation}
and
\begin{equation}
    \Delta h_{\times}(t;e_0,T_{\rm obs})
    =
    h_{\times,{\rm DM}}(t;e_0,T_{\rm obs})
    -
    h_{\times,{\rm disk}}(t;e_0,T_{\rm obs}) .
\end{equation}

The residual signal-to-noise ratio is denoted by $\rho_{\Delta}$. In the
present calculation, it is evaluated as
\begin{equation}
    \rho_{\Delta}^{2}
    =
    4
    \int_{f_{\min}}^{f_{\max}}
    \frac{
    \left|
    \Delta \tilde h_{+}(f)
    \right|^2
    +
    \left|
    \Delta \tilde h_{\times}(f)
    \right|^2
    }{
    S_n^{\rm LISA}(f)
    }\,df .
    \label{eq:rho_delta}
\end{equation}
Here $S_n^{\rm LISA}(f)$ is the LISA noise power spectral density, for which
we use the analytic sensitivity curve given in
Ref.~\cite{RobsonCornishLiu2019}.
When $\rho_{\Delta}\lesssim 1$, the DM--disk residual is comparable to or
smaller than the LISA noise level, indicating that the two effects are
difficult to distinguish. When $\rho_{\Delta}\gtrsim 1$, the residual becomes
detectable in a noise-weighted sense, while $\rho_{\Delta}\gg 1$ indicates
significant DM--disk waveform distinguishability within the LISA sensitivity
band~\cite{CutlerFlanagan1994,JaranowskiKrolak2012,LindblomOwenBrown2008}.

We adopt
the instrumental LISA noise spectrum~\cite{RobsonCornishLiu2019}
\begin{equation}
    S_n^{\rm LISA}(f)
    =
    \frac{10}{3L^2}
    \left[
    P_{\rm OMS}(f)
    +
    2\left(1+\cos^2\frac{f}{f_*}\right)
    \frac{P_{\rm acc}(f)}{(2\pi f)^4}
    \right]
    \left[
    1+\frac{6}{10}\left(\frac{f}{f_*}\right)^2
    \right],
    \label{eq:lisa_psd}
\end{equation}
where
\begin{equation}
    f_*=\frac{c}{2\pi L},
    \qquad
    L=2.5\times 10^9\,{\rm m}.
\end{equation}

The optical metrology noise and acceleration noise are given by
\begin{equation}
    P_{\rm OMS}(f)
    =
    \left(1.5\times10^{-11}\,{\rm m}\right)^2
    \left[
    1+
    \left(
    \frac{2\times10^{-3}\,{\rm Hz}}{f}
    \right)^4
    \right]
    {\rm Hz}^{-1},
\end{equation}
and
\begin{equation}
    P_{\rm acc}(f)
    =
    \left(3\times10^{-15}\,{\rm m\,s^{-2}}\right)^2
    \left[
    1+
    \left(
    \frac{0.4\times10^{-3}\,{\rm Hz}}{f}
    \right)^2
    \right]
    \left[
    1+
    \left(
    \frac{f}{8\times10^{-3}\,{\rm Hz}}
    \right)^4
    \right]
    {\rm Hz}^{-1}.
\end{equation}

\begin{figure}[htbp]
    \centering
    \includegraphics[width=0.86\linewidth]{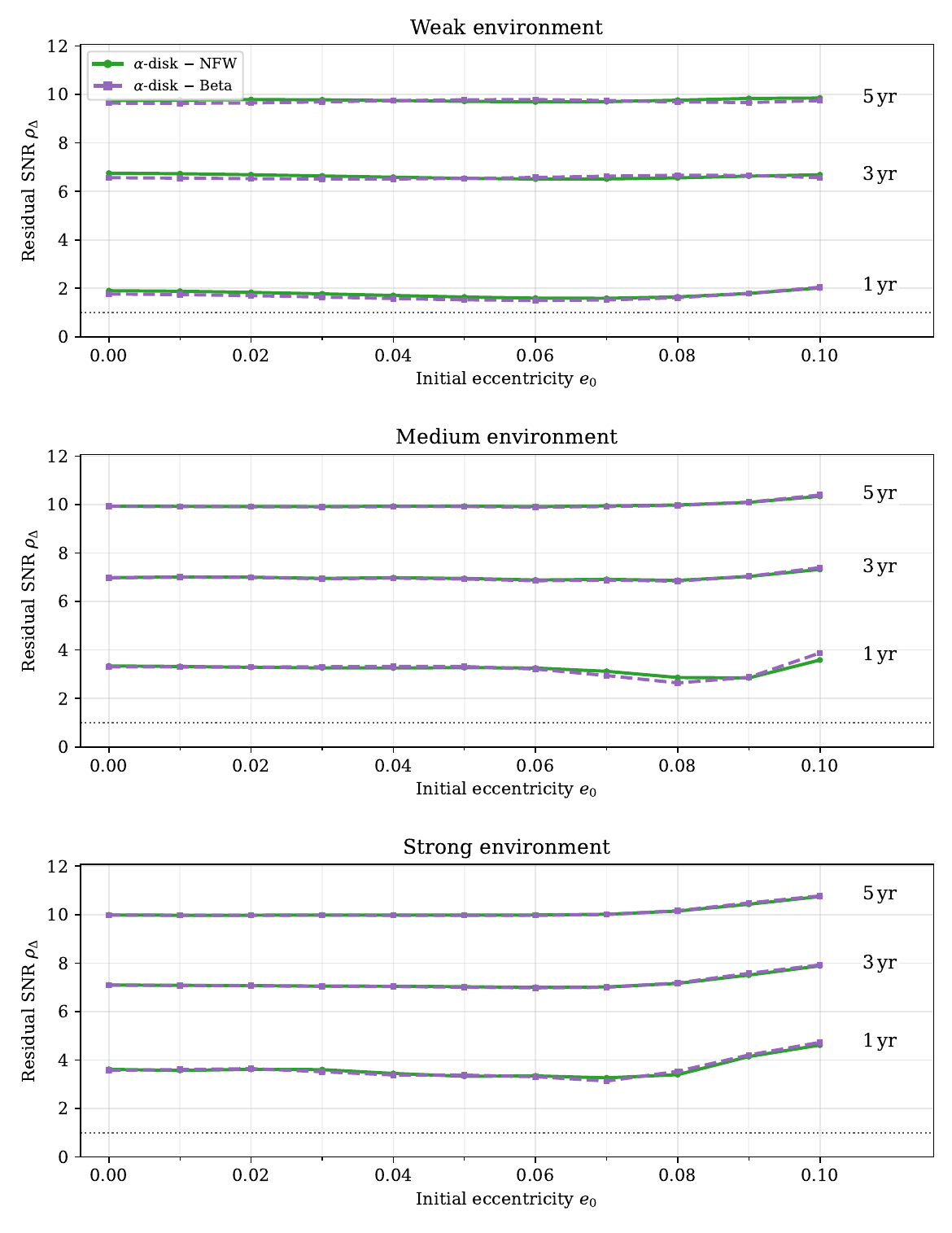}
    \caption{Residual SNR $\rho_{\Delta}$ between the accretion-disk and dark-matter
induced EMRI waveforms as a function of the initial eccentricity $e_0$
under weak, medium, and strong environmental settings. The results are shown
for $T_{\rm obs}=1,3,5\,{\rm yr}$, with solid green and dashed purple curves
corresponding to the $\alpha$-disk--NFW and $\alpha$-disk--Beta cases,
respectively. The horizontal dotted line denotes the threshold
$\rho_{\Delta}=1$.}
    \label{fig:fig4}
\end{figure}

Using the residual-SNR expression in
\hyperref[eq:rho_delta]{(\ref*{eq:rho_delta})},
we compute the residual signal-to-noise ratio $\rho_{\Delta}$ between the
accretion-disk and dark-matter induced waveforms under the weak, medium, and
strong environmental settings described above. For each environment, we
further consider three observation times,
$T_{\rm obs}=1,3,5\,{\rm yr}$, and examine the dependence of
$\rho_{\Delta}$ on the initial eccentricity $e_0$, as shown in
Fig.~\hyperref[fig:fig4]{\ref*{fig:fig4}}.

The results indicate that the DM--disk waveform residual is distinguishable
for all the parameter configurations considered here, since
$\rho_{\Delta}$ remains above the reference threshold $\rho_{\Delta}=1$.
The distinguishability is mainly controlled by the observation time: longer
observations lead to systematically larger residual SNRs, showing that the
separation between the two environmental effects is primarily enhanced by
long-term phase accumulation.

In contrast, the dependence on the initial eccentricity is relatively weak
for $0\leq e_0\leq 0.1$. The residual SNR varies only mildly with $e_0$ in
the weak and medium environments. A more visible eccentricity dependence
appears in the strong-environment case, where $\rho_{\Delta}$ increases at
larger $e_0$, especially for the one-year observation. This suggests that
eccentricity can amplify the DM--disk waveform difference when the
environmental effect is sufficiently strong, but it is not the dominant
factor compared with the observation duration.

The nearly overlapping $\alpha$-disk--NFW and $\alpha$-disk--Beta curves
further show that, within the present parameter setup, the residual SNR is
not strongly sensitive to the choice of the dark-matter density profile.
These results indicate that the main factors governing the DM--disk distinguishability are
the observation time and the environmental strength, while the initial
eccentricity and the specific dark-matter profile play secondary roles.

\section{Conclusion}

In this work, we first constructed Numerical Kludge waveforms including both
dark-matter environmental effects and accretion-disk environmental effects.
Based on this waveform model, we introduced the normalized eccentricity
response function $S_{\rm env}(e_0)$ to quantify the sensitivity of different
environmental effects to the initial orbital eccentricity $e_0$. Under the
weak, medium, and strong environmental configurations considered here, the
phase dephasing induced by the accretion disk exhibits a much stronger
dependence on $e_0$ than that induced by the DM halo. This indicates
that the eccentricity response of the phase evolution can serve as an
effective diagnostic for distinguishing accretion-disk effects from
dark-matter effects.

We then investigated the phase distinguishability between the dark-matter and
accretion-disk environmental effects. Taking an accumulated phase difference
of $1\,{\rm rad}$ as the distinguishability threshold, we used the earliest
time at which the accumulated phase difference between the dark-matter and
accretion-disk waveforms reaches this threshold to characterize the
distinguishability time $T_{\rm dis}(e_0)$. This quantity allows us to examine
how different initial eccentricities affect the time required to distinguish
the two environmental effects.

For all three environmental strengths, the distinguishability time exhibits a
smooth and weakly non-monotonic dependence on the initial eccentricity
$e_0$. Although eccentricity provides a moderate modulation of the phase
distinguishability, the environmental strength plays a more important role in
determining the observation time required to distinguish the two environmental
effects.

To connect the phase-level distinguishability with detector sensitivity, we
further evaluated the residual signal-to-noise ratio between the
dark-matter-induced and accretion-disk-induced waveforms using the LISA noise
power spectral density. For the benchmark parameter choices adopted in this
work, the residual SNR remains above the reference threshold
$\rho_\Delta=1$, indicating that the waveform difference between the two
environmental models can be detected in a noise-weighted sense. The residual
SNR increases systematically with the observation time, confirming that
long-term phase accumulation is the dominant factor controlling the DM--disk
distinguishability. By comparison, its dependence on the initial eccentricity
is relatively weak in the weak and medium environments, although it becomes
more visible in the strong-environment case.

Taken together, these findings suggest that the accretion-disk effect is more
sensitive to the initial eccentricity $e_0$ than the dark-matter environmental
effect. Therefore, the $e_0$-dependent phase response can be used as a useful
auxiliary diagnostic for distinguishing accretion-disk effects from
dark-matter effects, or from other environmental effects that are relatively
insensitive to $e_0$. However, the LISA residual signal-to-noise ratio analysis
also shows that, in a noise-weighted detection sense, the dominant factor
controlling the distinguishability between the two waveforms is still the
observation duration. As the observation time increases, the accumulated phase
difference continues to grow, leading to a systematic increase in the residual
SNR. The role of the initial eccentricity is therefore mainly to modulate the
environmental phase response and provide auxiliary discriminatory information,
rather than to act as the primary factor determining detectability.
\setcounter{figure}{0}
\renewcommand{\thefigure}{A\arabic{figure}}
\appendix

\section{Comparison between Newtonian-order and relativistic dark-matter metrics}
\label{app}

In the main text, the DM halo effect is incorporated through the modified
metric function introduced in Eq.~%
\eqref{eq:f_dm_general}%
. This treatment is obtained from the Newtonian circular velocity relation
and therefore corresponds to the leading-order contribution of the DM
gravitational potential.

In the present model, the metric takes a Schwarzschild-like form with a
single metric function, namely
\begin{equation}
ds^2=-F(r)dt^2+\frac{dr^2}{F(r)}
+r^2d\Omega^2.
\end{equation}

In contrast, Ref.~%
\cite{Zhao:2023tyo}%
constructs the DM-modified spacetime by solving the Einstein equations.
The corresponding metric is written in a more general form.

\begin{equation}
ds^2=-A(r)dt^2+\frac{dr^2}{B(r)}
+r^2d\Omega^2.
\end{equation}

In this relativistic construction, the two metric functions are determined
independently by the DM energy distribution, leading generally to

\begin{equation}
\displaystyle
A(r)\neq B(r).
\end{equation}

The main difference between the two approaches is that the
relativistic solution includes higher-order curvature corrections associated
with the DM spacetime, whereas the present work retains only the
Newtonian-order correction to the gravitational potential.

In the weak-field limit, the relativistic relation reduces to

\begin{equation}
\frac{d\ln\sqrt{A(r)}}{dr}
\simeq
\frac{m(r)}{r^2}
,
\end{equation}
which recovers the Newtonian relation used to construct
$f_{\rm DM}(r)$
at leading order.

Thus, the metric adopted in this work captures the leading-order
weak-field relation of the relativistic DM construction. The main neglected effects are
the higher-order relativistic corrections and the difference between the
temporal and radial metric components. These corrections may modify the
orbital frequencies and relativistic precession, resulting in additional
phase corrections in long-duration EMRI waveforms.

The magnitude of these effects depends on the DM density profile, halo
normalization, and orbital configuration. The relativistic construction
provides a more self-consistent description in the strong-field region,
while the present model captures the dominant DM-induced environmental
effect considered in this work. To examine whether adopting the relativistic DM treatment affects our main
conclusions, we repeat the eccentricity-dependence analysis using the
relativistic DM configurations considered in Ref.
\cite{Zhao:2023tyo}.
We adopt the eccentricity response function
$S_{\rm env}(e_0)$
defined in Eq.~%
\eqref{eq:eccentricity_sensitivity}%
, and keep the remaining waveform parameters unchanged from
Table~%
\ref{tab:env_parameters}%
.

\begin{figure}[htbp]
\centering
\includegraphics[width=0.86\linewidth]{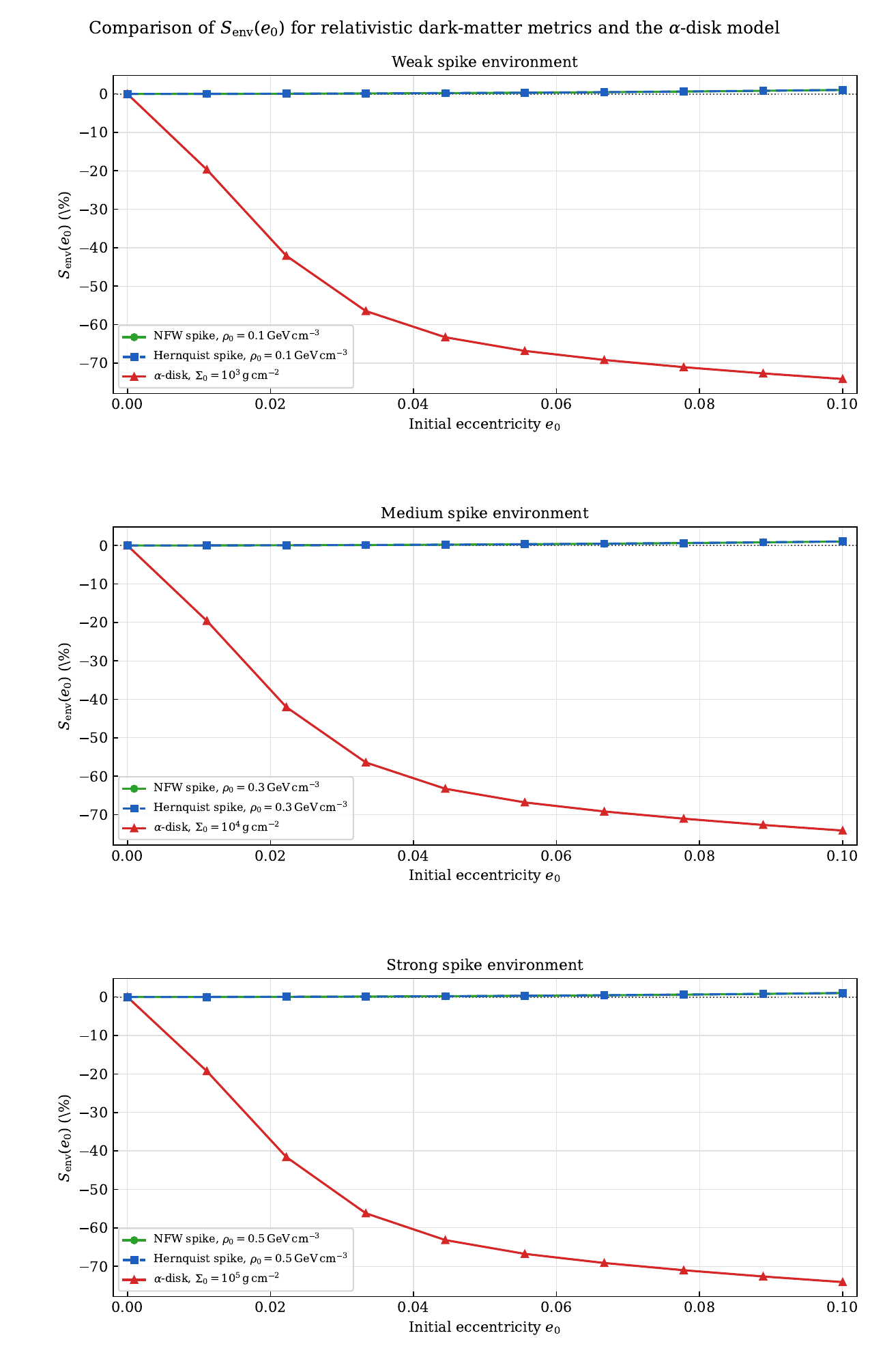}
\caption{Normalized eccentricity response $S_{\rm env}(e_0)$ obtained using the
relativistic dark-matter metric.}
\label{fig:relDM}
\end{figure}

For the relativistic DM configurations, the adopted DM model and the choices of
the weak, medium, and strong environmental parameters are based on the
parameters considered in Ref.
\cite{Zhao:2023tyo}%
, with the NFW spike and Hernquist spike models and DM densities
$\rho_{\rm DM}=0.1,\ 0.3,\ 0.5~{\rm GeV\,cm^{-3}}$
for the weak, medium, and strong configurations, respectively. The corresponding results are shown in Fig.~%
\ref{fig:relDM}%
.

Although the relativistic metric leads to quantitative changes in the
accumulated phase shift, the qualitative behavior remains unchanged. The
DM-induced dephasing is still weakly dependent on the initial eccentricity,
while the accretion-disk-induced dephasing shows a much stronger dependence
on $e_0$. Therefore, adopting the relativistic DM configurations does not qualitatively change the main conclusions presented in the main text.
\setcounter{figure}{0}
\renewcommand{\thefigure}{B\arabic{figure}}
\section{Additional test with the Coulomb logarithm $\ln\Lambda\simeq10$}
\label{app:lambda10}

\begin{figure}[htbp]
\centering
\includegraphics[width=0.86\linewidth]{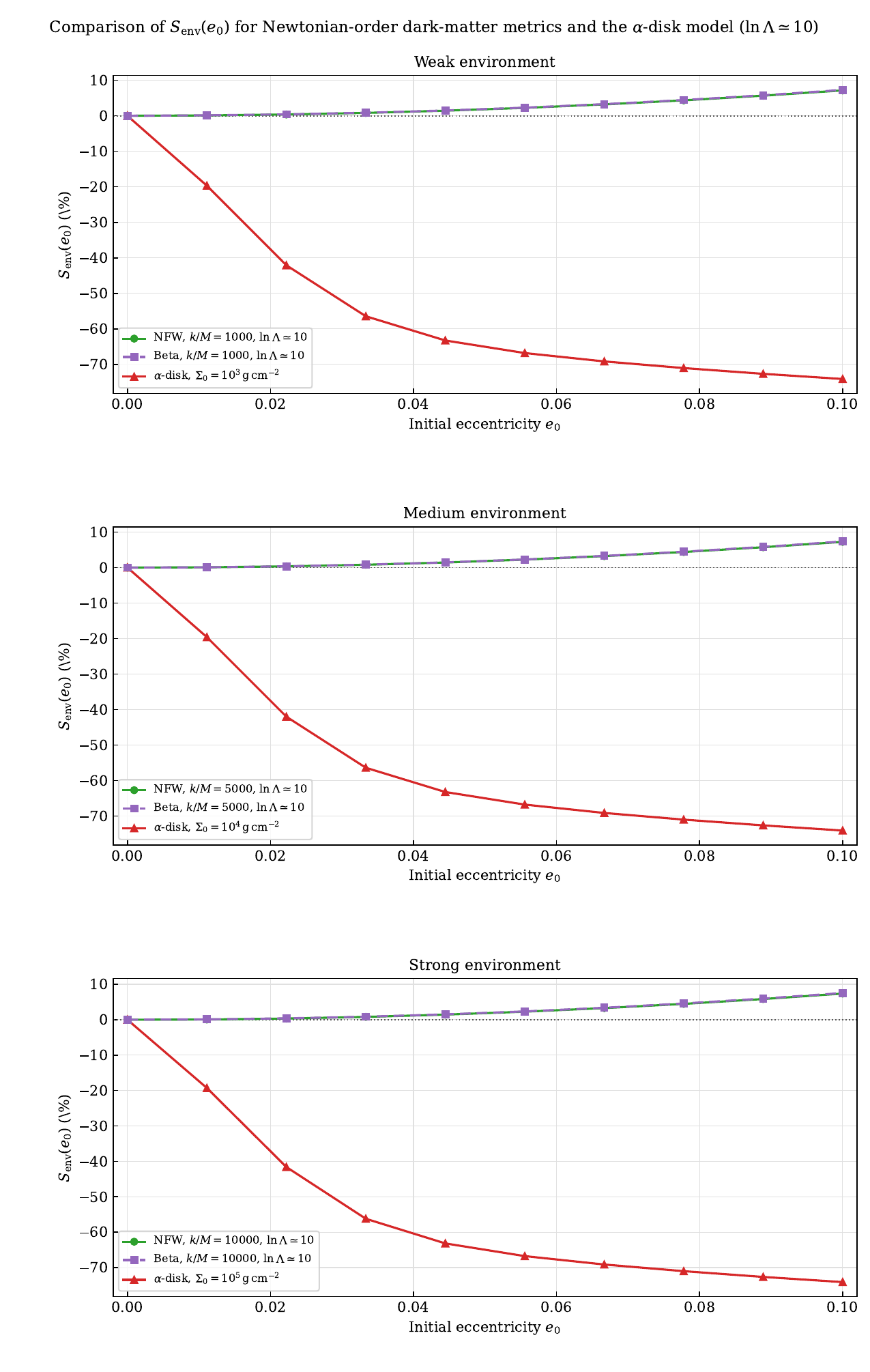}
\caption{The eccentricity sensitivity $S_{\rm env}(e_0)$ obtained with
$\ln\Lambda\simeq10$ using the original dark-matter metric.}
\label{fig:lambda10_old}
\end{figure}

We additionally consider the case
$\ln\Lambda\simeq10$\cite{Yue:2019}. Using the eccentricity sensitivity
$S_{\rm env}(e_0)$
defined in Eq.
\eqref{eq:eccentricity_sensitivity}
, we repeat the calculation for the Newtonian-order dark-matter
metric model adopted in this work, while keeping the remaining parameters
the same as those listed in Table
\ref{tab:env_parameters}. The resulting eccentricity sensitivity curves are shown
in Fig.
\ref{fig:lambda10_old}
.

The results demonstrate that changing the Coulomb logarithm from
$\ln\Lambda\simeq3$
to
$\ln\Lambda\simeq10$
modifies the quantitative strength of the dynamical-friction contribution.
However, the overall behavior of
$S_{\rm env}(e_0)$
and its dependence on the initial eccentricity remain unchanged.
Therefore, the main conclusions regarding the eccentricity sensitivity of
the environmental effects are robust against the choice of the Coulomb
logarithm.

\section*{Acknowledgements}
This work was supported by the National Natural Science Foundation of China under Grant No. 12065007.

\bibliographystyle{apsrev4-1}
\bibliography{refs}

\end{document}